\documentclass[aps,prc,twocolumn,superscriptaddress,showpacs,
floatfix]{revtex4-1}
\usepackage{amsmath, amsthm, amssymb}
\usepackage{graphicx,subfigure}
\usepackage{comment}
\usepackage{bm}
\usepackage{color}

\newcommand{\vecr}{{\bm r}}
\newcommand{\vecR}{{\bm R}}

\newcommand{\vecK}{{\bm K}}

\def\nuc#1#2{\relax\ifmmode{}^{#1}{\protect\text{#2}}\else${}^{#1}$#2\fi}

\newcommand {\beq} {\begin{eqnarray}}
\newcommand {\eeq} {\end{eqnarray}}
\newcommand {\eeqn} [1] {\label{#1} \end{eqnarray}}%

\begin{document}

\title{Rapid convergence of the Weinberg expansion of the deuteron stripping
amplitude}
\author{D. Y. Pang}
\affiliation{School of Physics and Nuclear Energy Engineering, Beihang
University, Beijing, 100191, China}
\author{N. K. Timofeyuk}
\affiliation{Department of Physics, Faculty of Engineering and Physical
Sciences,
University of Surrey, Guildford, Surrey GU2 7XH, United Kingdom}
\author{R. C. Johnson}
\affiliation{Department of Physics, Faculty of Engineering and Physical
Sciences,
University of Surrey, Guildford, Surrey GU2 7XH, United Kingdom}
\author{J. A. Tostevin}
\affiliation{Department of Physics, Faculty of Engineering and Physical
Sciences,
University of Surrey, Guildford, Surrey GU2 7XH, United Kingdom}

\date{\today}

\begin{abstract}
Theories of  $(d,p)$ reactions frequently use a formalism based on a transition
amplitude that is dominated by the components of the total three-body
scattering
wave function where the spatial separation between the incoming neutron and
proton is confined by the range of the $n$-$p$ interaction, $V_{np}$. By
comparison with calculations based on the continuum discretized coupled
channels method we show that the $(d,p)$
transition amplitude is dominated by the first term of the expansion of the
three-body wave function in a complete set of Weinberg states. We use the
\nuc{132}{Sn}$(d,p)$\nuc{133} {Sn} reaction at 30 and 100 MeV as examples of
contemporary interest. The generality of this observed dominance and its
implications for future theoretical developments are discussed.
\end{abstract}


\maketitle

\section{Introduction}
There is growing interest and activity in transfer reaction studies using
radioactive beams, driven by increased secondary beam intensities and motivated
by the search for new physics at the edge of nuclear stability
\cite{Jonson-PR-2004,Cat10} and by the need for low-energy reaction rates for
astrophysical applications \cite{Tom07,Mukhamedzhanov-PRC-2008}. The $(d,p)$
reaction, measured in inverse kinematics, is well suited for these purposes. It
can provide spin-parity assignments for nuclear states, allow determination of
spectroscopic
strengths of single-particle configurations, and give asymptotic normalization
coefficients in the tail of overlap functions. The reliability of this deduced
nuclear structure information depends on the existence of a reaction theory
that
describes adequately the mechanism of the $(d,p)$ reaction.

This paper uses a formulation of the $A(d,p)B$ reaction amplitude that
emphasizes the components of the total neutron+proton+target scattering wave
function where the spatial separations between the incoming neutron and proton
are confined by the range of the $n$-$p$ interaction, $V_{np}$. These
components
contain both the bound and continuum states of the $n$-$p$ system. Since the
$n$-$p$ binding energy in the deuteron is small and the optical potentials that
generate the tidal break-up forces are smooth functions of position, the
strength of inelastic excitations to the $n$-$p$ continuum is expected to be
concentrated at low $n$-$p$ relative energies. This suggests that the coupling
effects between different $n$-$p$ states can be treated adiabatically and leads
to a simple prescription for calculating the scattering wave function at small
$n$-$p$ separations \cite{Johnson-PRC-1970}.

In the adiabatic model the $A(d,p)B$ transition amplitude has exactly the same
structure as that of the distorted-wave Born approximation (DWBA), for which
many computer codes are available, which has led to its widespread use
\cite{Johnson-PRC-1970,Harvey-PRC-71, Wales-NPA-76,Johnson-NPA-1974,
Jenny-PRL-2010, Cat10}. The adiabatic model frequently provides significant
improvements over the DWBA for $A(d,p)B$ angular distributions and giving
consistent results for nuclear structure information \cite{Jenny-PRL-2010}.

There are two key ingredients in the adiabatic model:\\
(i) the assumption that only components of the three-body scattering wave
function with small $n$-$p$ separation are needed for the $A(d,p)B$ transition
amplitude and
(ii) the validity of the adiabatic treatment of deuteron break-up at the
nuclear surface.
The primary purpose of this paper is to show that assumption
(i) is justified for a useful range of reaction energies when it
is implemented in terms of a precisely defined projection of the three-body
scattering wave function. This projection will be shown
to involve the first Weinberg state component of
the full wave function.

Investigations of how assumption (ii) influences the predicted $(d,p)$ cross
sections were carried out using the quasi-adiabatic model
\cite{Amakawa-PRC-1984, Stephenson-PRC-1990}, the Weinberg states expansion
(WSE) method \cite{Johnson-NPA-1974,Laid-PRC-1993}, the continuum discretized
coupled channels (CDCC) method, and also Faddeev equation methods
\cite{Deltuva-PRC-2007,Filomena-PRC-2012}. The importance of nonadiabatic
effects has been found to depend on the target and incident energy and, in the
worst cases, these affected both the shapes and the magnitudes of the
calculated
differential cross sections \cite{Laid-PRC-1993, Filomena-PRC-2012}. There is
therefore an important need to provide a practical way of introducing
corrections to the adiabatic approximation. Our aim here is to provide a
suitable definition of the projection of the full scattering wave function
implied by assumption (i), which we call the first Weinberg projection. We show
that this projection, which is a function of only a single vector coordinate,
dominates the calculation of the $A(d,p)B$ transition amplitude. This result
implies that to include effects beyond the adiabatic approximation one can
focus
on improvements to the calculation of this projection only.

The CDCC method for solving the three-body problem does not use the adiabatic
approximation (ii). From a practical point of view it is well adapted to the
study of deuteron breakup effects on $A(d,p)B$ reactions. In principle in the
CDCC
method one attempts to calculate the three-body scattering wave function in the
whole six-dimensional coordinate space of the neutron+proton+target ($n+p+A$)
three-body system. Our approach is to compare calculations of the $(d,p)$
transition amplitude made using a complete CDCC wave function with calculations
which retain only the first few Weinberg components of the full CDCC wave
function.

In Sec. II we describe how the projection procedure mentioned above is related
to the Weinberg state and CDCC expansion methods and we connect these. In Sec.
III we construct the Weinberg components using the CDCC wave functions and
in Sec. IV we compare calculations of the $(d,p)$ transition amplitudes using
the first few Weinberg components. We summarize our results in Sec. V.

\section{Three-body wave function and its expansion in the CDCC and Weinberg
state bases} \label{sec-expansion}
In the absence of inelastic excitations of the target and residual nuclei $A$
and $B$ in the incident and outgoing channels, the transition amplitude of
the $A(d,p)B$ reaction can be written as \cite{Johnson-PRC-1970}
\begin{equation}\label{exact}
 T_{dp} = \langle \chi_p^{(-)}I_{AB}|V_{np}|\Psi^{(+)}\rangle\,.
\end{equation}
Here $\chi_p^{(-)}$ is the outgoing proton distorted wave (where we neglect
certain
$1/A$ corrections \cite{JT}), $I_{AB}$ is the overlap function between the
wave functions of $A$ and $B$, $V_{np}$ is the neutron-proton interaction,
and $\Psi^{(+)}$ is the projection of the full many-body wave function onto
the three-body, $n+p+A$, channel with $A$ in its ground state. The effect of
coupling to excited states of $A$ is implicitly taken into account through
the use of complex nucleon optical potentials, but contributions from
transitions that explicitly excite components of $A$ in the initial state
and $B$ in the final state are ignored. We assume that $\Psi^{(+)}$ satisfies
the Schr\"{o}dinger equation
\begin{eqnarray}\label{the-Schrodinger-eq}
\left[E_d+i\epsilon -H_{np}-T_R-U_n(\vecr_n)-U_p(\vecr_p)\right]
\Psi^{(+)}(\vecr,\vecR)\nonumber\\
= i\epsilon\phi_d(\vecr)e^{i\vecK_d\cdot\vecR},&&
\end{eqnarray}
where $H_{np}=T_r+V_{np}$ is the $n$-$p$ relative motion Hamiltonian. Here
$E_d=E_\textrm{c.m.}-\epsilon_d$ where $\epsilon_d$ is the deuteron binding
energy and $E_\textrm{c.m.}$ is the three-body energy in the center-of-mass
system. $U_n$ and $U_p$ are the optical model potentials for the neutron and the
proton with the target nucleus, respectively, and $\vecK_d$ is the wave number
associated with $E_d$. The coordinates $\vecr_p$ and $\vecr_n$ are the proton
and neutron coordinates with respect to the target $A$ while $\vecr=\vecr_p-
\vecr_n$ and $\vecR=\tfrac{1}{2}(\vecr_n+\vecr_p)$ are the relative and c.m.
coordinates of the $n$-$p$ pair. Also,
$$
T_r=-\frac{\hbar^2}{2\mu_{np}}\nabla_r^2\ \textrm{ and }
\ T_R=-\frac{\hbar^2}{2\mu_{dA}}\nabla_R^2
$$
are the kinetic energy operators associated with $\vecr$ and $\vecR$, with
$\mu_{np}$ and $\mu_{dA}$ the reduced masses of the $n$-$p$ pair and the
$n+p+A$ system, respectively. The right hand side of Eq.
(\ref{the-Schrodinger-eq}) specifies the incident boundary condition of a
deuteron with initial wave function $\phi_d$ and the physical total wave
function is to be calculated in the limit $\epsilon\rightarrow 0+$. The
superscripts on $\chi_p^{(-)}$ and $\Psi^{(+)}$ indicate that they obey
ingoing and outgoing waves boundary conditions, respectively. For simplicity,
these superscripts are omitted in the following text.

In the next two sections we describe two expansion schemes for the
total wave function $\Psi(\vecr,\vecR)$.

\subsection{The Weinberg states expansion}
For $n$-$p$ separations $\vecr$ within the range of $V_{np}$, the wave function
$\Psi(\vecr,\vecR)$ has the expansion \cite{Johnson-NPA-1974, Laid-PRC-1993}
\begin{equation}\label{eq-expansion-wse}
\Psi(\vecr,\vecR) = \sum_i\phi_i^{W}(\vecr)\chi_i^{W}(\vecR),
\end{equation}
where the Weinberg states, $\phi_i^{W}$, are solutions of the equation
\begin{equation}\label{weinberg-states}
[-\epsilon_d-T_r-\alpha_i V_{np}]\phi_i^{W}(\vecr)=0, \ \ i=1,2,\ldots
\end{equation}
with fixed energy $-\epsilon_d$ and eigenvalues $\alpha_i$. For radii
$r>r_i$, where $r_i$ is such that $\alpha_i V_{np}(r)$ is negligible,
all of the Weinberg states decay exponentially, like the deuteron ground
state wave function. For $r<r_i$, they oscillate with a wavelength that
varies with $i$, becoming increasingly oscillatory with increasing $i$
(see the examples given in \cite{Laid-PRC-1993} for the case of a
Hulth\'en form for $V_{np}$).

The Weinberg states form a complete set of functions of $r$ for regions of
the $r$ axis on which $V_{np}$ is non-vanishing. They are therefore well
adapted to expanding $\Psi$ in this region. They do not satisfy the usual
orthonormality relation but instead satisfy
\begin{equation}
\langle\phi_i^{W}|V_{np}|\phi_j^{W}\rangle=-\delta_{ij}\,,\label{Worthog}
\end{equation}
where the value $-1$ for $i=j$ has been chosen for convenience.

This form of orthonormality, with a weight factor $V_{np}$, means that if one
wishes to represent an arbitrary state $\varphi(r)$ as a linear superposition
of
Weinberg states then the unique choice of coefficients $a_i$ which minimizes
the
difference
\begin{equation}
\Delta = \int d\vecr\,V_{np}\mid \varphi-\sum_i a_i \phi_i^{W}\mid^2\,,
\label{diff}
\end{equation}
is
\begin{equation}
a_i= -\langle\phi_i^{W}|V_{np}|\varphi \rangle \label{ai}\,.
\end{equation}
Use of a factor $V_{np}$ in Eq. (\ref{diff}), which weights $r$ values
according
to $V_{np}$, provides a natural scheme for constructing the expansion
coefficients for states of $n$-$p$ relative motion for use in the $(d,p)$
transition amplitude.

\subsection{The CDCC basis method}
The CDCC method involves the expansion of $\Psi(\vecr,\vecR)$ in terms of a
complete set of $n$-$p$ continuum bin states $\phi_i^{bin}$  (see, e.g., Ref.
\cite{Austern-PR-1987}), written
\begin{equation}\label{eq-expansion-cdcc}
\Psi(\vecr, \vecR) = \phi_d(\vecr)\chi_0(\vecR)+\sum_{i=1}
\phi_i^{bin}(\vecr)\chi_i^{bin}(\vecR)\,.
\end{equation}
The bin states are linear superpositions of continuum eigenfunctions of
$H_{np}$, on chosen intervals $\Delta k_i$ of $n$-$p$ continuum wave numbers,
and are orthogonal in the usual sense. So, the projection of the three-body
Schr\"{o}dinger equation of Eq. (\ref{the-Schrodinger-eq}) onto this set of
spatially-extended bin states leads to a set of coupled-channel equations
for the channel wave functions $\chi_i^{bin}(\vecR)$. The coupling potentials,
generated from the nucleon optical potentials, are long-ranged and link
parts of the wave function from all $n$-$p$, $n$-$A$, and $p$-$A$ separations.

These CDCC equations can be solved numerically and their convergence
properties have been intensively studied.

\subsection{Connection between the CDCC and Weinberg basis wave functions}
It is known from experience with CDCC calculations that the energy range of
$n$-$p$ continuum states that are coupled to the incident deuteron channel is
limited to tens of MeV. Thus, we expect that inside the range of $V_{np}$
the wave function $\Psi(\vecr,\vecR)$ will not be a strongly oscillatory
function of $r$ and only a few terms of the Weinberg expansion will be needed
to evaluate the $(d,p)$ matrix element. Note that this has nothing to do with
the strength of the coupling between Weinberg components in $\Psi(\vecr,\vecR)$
or how rapidly the Weinberg expansion for $\Psi(\vecr,\vecR)$ itself converges,
but rather it relates to how rapid the convergence of the sequence of
contributions to the
$(d,p)$ amplitude is from the different Weinberg components. We do not obtain
the
latter from a set of coupled equations, as, e.g., was done successfully in Ref.
\cite{Laid-PRC-1993}, but rather from the CDCC expansion of $\Psi(\vecr,\vecR)$.
The quantitative issues arising from a comparison with the approach of Ref.
\cite{Laid-PRC-1993} will be addressed elsewhere.

To connect the Weinberg and CDCC components of $\Psi(\vecr,\vecR)$ we project
$\Psi$, expressed in the CDCC basis, onto individual Weinberg states using the
orthogonality property of Eq. (\ref{Worthog}). The Weinberg distorted waves,
$\chi_i^{W}$, and those of the CDCC basis, $\chi_j^{bin}$, are related using
\begin{equation}\label{projection}
\chi_{i}^{W}(\vecR)=C_{i0}\chi_0(\vecR) + \sum_{j=1}C_{ij}\chi_j^{bin}(\vecR).
\end{equation}
The transformation coefficients $C_{ij}$ are given by
\begin{eqnarray}
C_{i0}&=&-\langle\phi_i^{W}|V_{np}|\phi_d\rangle,\ \ (=0,\,\,i\neq
1)\,,\nonumber \\
C_{ij}&=&-\langle\phi_i^{W}|V_{np}|\phi_j^{bin}\rangle, (i,j=1,2,\ldots)\,.
\label{Cij}
\end{eqnarray}

These coefficients also appear in the formulas
\begin{eqnarray}
\mid \phi_j^{bin}\rangle=\sum_i C_{ij}\mid \phi_i^W\rangle\,,\label{Cij2}
\end{eqnarray}
and
\begin{eqnarray}
\int d\vecr V_{np} \mid\phi_j^{bin}(r)\mid^2=\sum_i\mid C_{ij}\mid^2
\label{Cij3}
\end{eqnarray}
that quantify the contribution of each Weinberg state to a particular CDCC
bin state, in the presence of the weight factor $V_{np}$.

These $C_{ij}$ are determined entirely by the bound and scattering states of
$V_{np}$ in the energy range of the relevant bin states. They do not depend
on any other details of the reaction, such as the deuteron incident energy,
the transferred angular momentum, or the structure of the target nuclei.
The values of $C_{ij}$ do depend on how the CDCC bin states were constructed,
the bin sizes $\Delta k_i$, etc., however we have checked that the changes in
the computed $\chi_i^W$ are less than 0.1\% with typical choices of bin
sizes, such as $\Delta k_i \approx 0.1$-$0.15$ fm$^{-1}$. Throughout this
work a Hulth\'{e}n potential was used for $V_{np}$, namely,
\begin{equation}
V_{np}(r)= V_0/(e^{\beta r}-1),
\end{equation}
with parameters $V_0=-84.86$ MeV and $\beta=1.22$ fm$^{-1}$
\cite{Laid-PRC-1993}.
Only $s$-wave continuum states were included. These give the largest
contribution
to $\Psi(\vecr, \vecR)$ at small $r$.

\begin{figure}[htbp]
\centering
\includegraphics[width=0.48\textwidth]{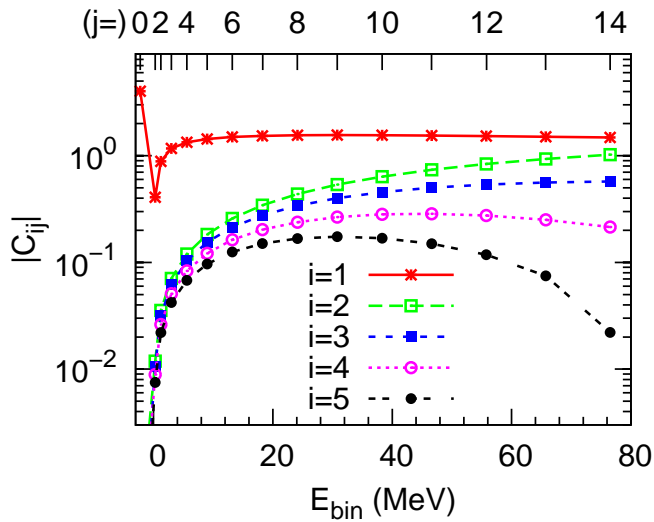}
\caption{(Color online) CDCC bin-state to Weinberg state transformation
coefficients $C_{ij}$, of Eq. (\ref{projection}), for Weinberg states
$i=1,2,\ldots,5$ and CDCC bin states $j=1,2,\ldots,14$. The deuteron
ground state is denoted by $j=0$. The CDCC bins were calculated up to
$n$-$p$ relative momenta $k_{max}=1.4$ fm$^{-1}$ in steps $\Delta k_i=0.1$
fm$^{-1}$. See Sec. \ref{sec-numerical-calc} for full details.}
\label{fig01}
\end{figure}

In Fig. \ref{fig01} we show the calculated $C_{ij}$ for $i\leq5$
for bin states $\phi_j^{bin}$ calculated from CDCC calculations using the
computer
code \textsc{fresco} \cite{fresco}. The lower and upper horizontal axes show the
$n$-$p$ continuum energies included in the CDCC and the label of the different
bins, with $j=1,\ldots,14$, respectively. The point with $j=0$ shows the
$C_{10}$
that connects with the deuteron ground state. Each line then corresponds to a
different Weinberg state, $\phi^W_i$.

For the $(d,p)$ reaction the most relevant continuum energies lie in the range
0 to 40 MeV. From Eq. (\ref{Cij2}) and the $i$ dependence of the $C_{ij}$ for
the lower energy (and $j$) bins in Fig. \ref{fig01}, we see that
the bin states in the relevant energy range are dominated by the first Weinberg
component with only small contributions from Weinberg states $i=2$-$5$. This
dominance is particularly marked for the low-energy continuum, which is the most
strongly coupled to the deuteron ground state by the break-up mechanism and that
has the largest $\chi_i^{bin}(\vecR)$ in Eq. (\ref{eq-expansion-cdcc}). At
the higher continuum energies the bin states are mixtures of several Weinberg
states, as was expected.

In Eq. (\ref{projection}), this dominance of the $i=1$ coefficients
for low continuum energies will make $\chi_1^W$ the dominant Weinberg distorted
wave provided the contributions from continuum bins with energies greater than
of order 30 MeV are not large. In the next section we present the details of
CDCC calculations and show that these qualitative observations are borne out
quantitatively for typical $(d,p)$ reactions and energies.

\section{Construction of the $\chi_i^{W}$ from the CDCC wave function}
\label{sec-numerical-calc}
In this section, as relevant topical examples, we construct the Weinberg
distorted waves $\chi_i^W$ for the $^{132}$Sn($d,p)^{133}$Sn reaction at
deuteron incident energies $E_d =$ 100 and 30 MeV at which the
contributions from closed channels are negligible. Neutron-rich target
nuclei, for inverse kinematics ($d,p)$ experiments at such energies per
nucleon, are available at several modern radioactive ion beam facilities
including RIKEN \cite{Kubo-RIPS}, GANIL \cite{Villari-SPRIL}, NSCL
\cite{Morrissey-NPA-1997}, FLNR at Dubna \cite{Rodin-ACCULINNA}, and IMP at
Lanzhou \cite{SunZY-RIBLL}.

\begin{figure}[htbp]
\centering
\includegraphics[width=0.48\textwidth]{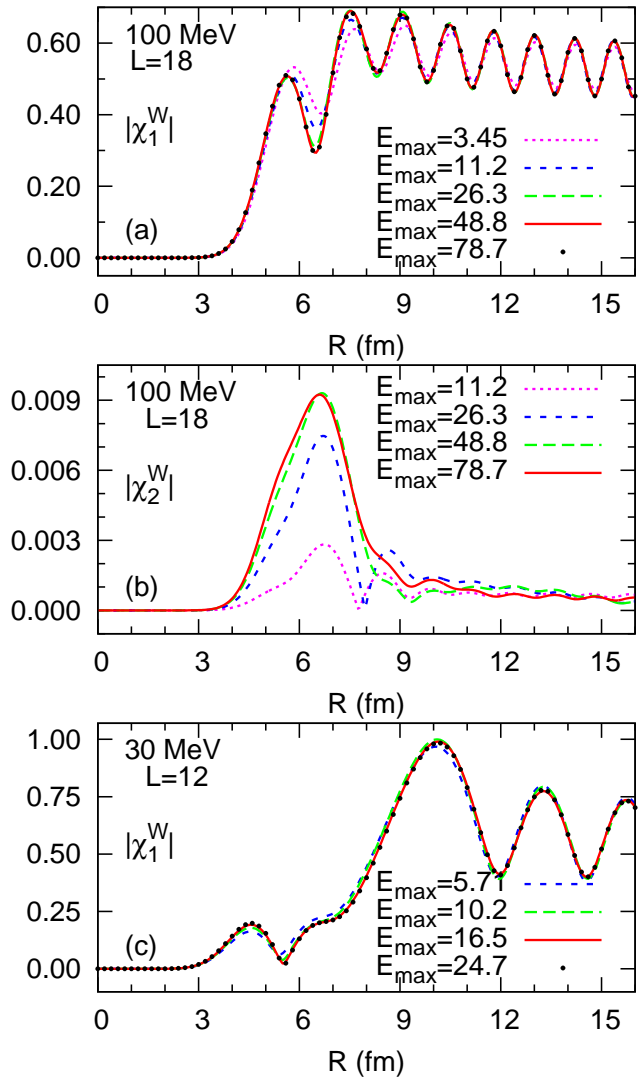}
\caption{(Color online) Convergence of selected partial waves of the Weinberg
components $\chi_i^{W}$ with respect to the maximum $n$-$p$ continuum energy
included in Eq.(\ref{projection}). Results are for (a) $\chi_1^{W}$ and $E_d=
100$ MeV, (b) $\chi_2^{W}$ and $E_d=100$ MeV, and (c) $\chi_1^{W}$ and $E_d=30$
MeV.
The partial wave values, $L$, associated with each $\chi_i^{W}$ are
indicated in each panel.} \label{fig02}
\end{figure}

We solved the CDCC equations using nucleon optical potentials, $U_n$ and $U_p$,
evaluated at half the incident deuteron energy, taken from the KD02 systematics
\cite{kd02}. Only the central parts of these potentials were used. Both the
nuclear and Coulomb potentials were used in constructing the coupling
potentials.
The continuum bin states $\phi^{bin}$ were computed by discretizing the $s$-wave
$n$-$p$ continuum using $\Delta k_i$ of 0.1 and 0.05 fm$^{-1}$ up to
$k_{max}=1.4$
and 0.75 fm$^{-1}$, corresponding to maximum continuum energies of 81.9 and 23.5
MeV, for $E_d = 100$ and 30 MeV, respectively. The coupled-channels CDCC
equations
were solved up to $R_{max} =100$ fm because of the long-range nature of the CDCC
couplings \cite{Jeff-PRC-2001}. The CDCC calculations were performed using the
computer code \textsc{fresco} \cite{fresco}.

\begin{figure}[htbp]
\centering
\includegraphics[width=0.48\textwidth]{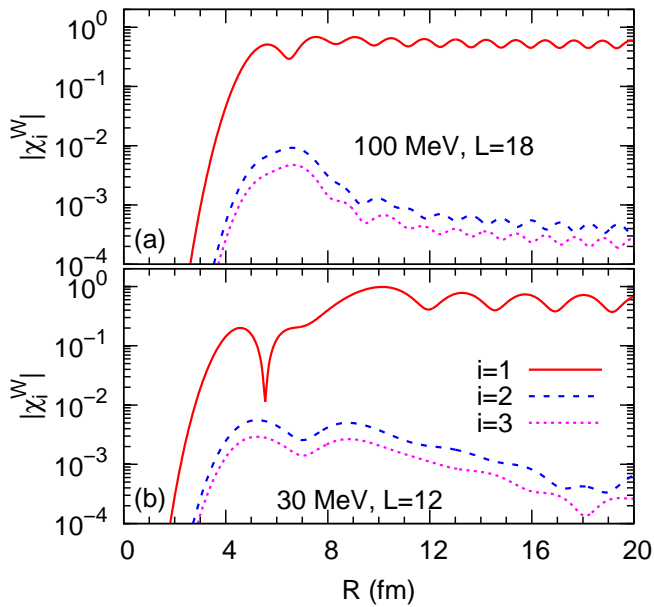}
\caption{(Color online) Calculated Weinberg state distorted waves $\chi_i^{W}$
demonstrating the dominance of $\chi_1^{W}$. Curves compare the moduli of
$\chi_1^{W}$, $\chi_2^{W}$ and $\chi_3^{W}$ for the $^{132}$Sn($d,p)^{133}$Sn
reaction for (a) $E_d= 100$ MeV and partial wave $L=18$, and (b) $E_d= 30$ MeV
and partial wave $L=12$.} \label{fig03}
\end{figure}

The $\chi^W_i$ were constructed from Eq. (\ref{eq-expansion-cdcc}) using the
coefficients $C_{ij}$ discussed in the previous section. It was found that bins
up to a maximum continuum energy of 25 MeV are sufficient for the convergence of
$\chi_1^W$ for both deuteron incident energies. This is illustrated in Figs.
2(a)
and 2(c), which show $\chi_1^W$ for partial waves with $L = 18$ and 12. Angular
momenta $L$ near these values drive the dominant contributions to the $(d,p)$
reaction cross sections for $E_d=100$ and 30 MeV, respectively. Convergence of
the $\chi_i^W$ with $i > 1$ was not achieved, as anticipated from the behavior
of the coefficients $C_{ij}$ shown in Fig. \ref{fig01}. We
demonstrate this in Fig. 2(b) for $\chi^W_2$ and $E_d = 100$ MeV. As is
expected, from the $C_{ij}$ dependence on $j$ for $i>1$, all $i>1$ Weinberg
components are about two orders of magnitude smaller than $\chi_1^W$ in the
most important radial region for the transfer amplitude. This is $R \approx 7$
fm in the present case (see Fig. \ref{fig03}).

\begin{figure}[htbp]
\centering
\includegraphics[width=0.48\textwidth]{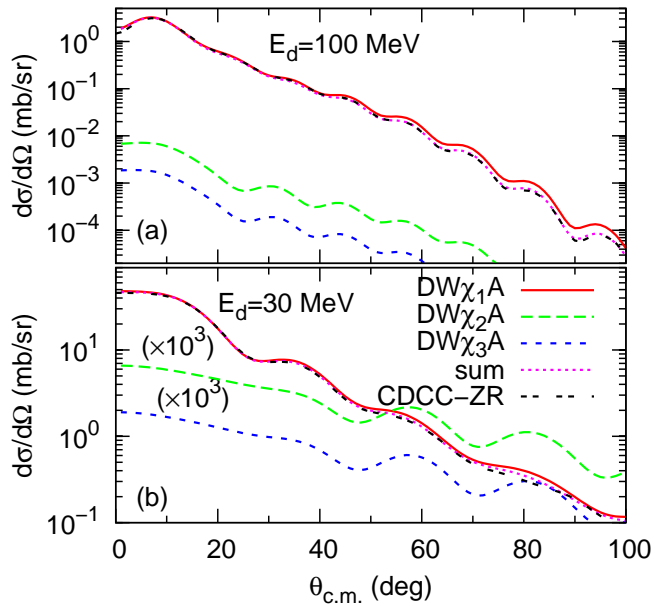}
\caption{(Color online) Comparisons of the calculated differential cross
sections for the \nuc{132}{Sn}$(d,p)$\nuc{133}{Sn} reaction at (a) 100 MeV
and (b) 30 MeV, using Weinberg distorted wave components $\chi_1^{W}$,
$\chi_2^{W}$, and $\chi_3^{W}$, showing the dominance of the first Weinberg
component $\chi_1^{W}$; see the text for details.}
\label{fig04}
\end{figure}

The Weinberg distorted wave components $\chi_i^W$ constructed above contain
contributions from all CDCC basis components within the range of $V_{np}$.
However, since $\chi_1^W$ dominates over all other Weinberg components, it is
sufficient to perform one-channel transfer reaction calculations with only
$\chi_1^W$ included. We call calculations truncated in this way DW$\chi_1$A
(distorted wave with $\chi_1^W$ approximation). For this purpose, we read the
calculated $\chi_i^W$ ($i=1,2,3$) into the computer code \textsc{twofnr}
\cite{twofnr} and calculate the transfer amplitude within the zero-range
approximation. We use the same KD02 optical potential systematics as used in the
deuteron channel for the proton distorted waves in the outgoing channel.
In the model calculations presented, the neutron overlap
function is approximated as a single particle wave function (with $\ell=3$)
calculated using a Woods-Saxon potential with standard radius and diffuseness
parameters, $r_0=1.25$ fm and $a_0=0.65$ fm and depth fitted to separation
energy of 2.47 MeV \cite{Audi-NPA-2003}. No spin-orbit potential was used for
this wave function.

The DW$\chi_i$A differential cross sections are shown
in Fig. \ref{fig04} for 100 and 30 MeV incident deuteron
energies, where the differential cross sections corresponding to each of
$\chi_{1,2,3}^W$ and their coherent sum are shown. Evident from these
figures is that the addition of channels $\chi_2^W$ and $\chi_3^W$  does not
influence the cross sections at the forward angles where the angular
distributions are usually measured and are most valuable for spectroscopy.
The $\chi_2^W$ and $\chi_3^W$ contributions are noticeable at large angles
where the cross sections are small, but even there the changes are small.
For comparison, the results of CDCC-ZR calculations, which include the
contributions to transfer (in the zero-range approximation) from all of the
CDCC continuum bins used to construct $\chi_i^W$, are also shown. As was
expected, the cross sections from the CDCC-ZR calculation and from the
coherent sums of the DW$\chi_i$A ($i=1,2,3$) amplitudes agree very well
at both of the energies studied.

\section{Summary}\label{summary}
Using as an example the $^{132}$Sn$(d,p)^{133}$Sn reaction at energies
of 15 and 50 MeV/nucleon
typical of modern radioactive ion beam facilities, we have demonstrated
that the dominant effects of deuteron breakup on calculations of $(d,p)$
reaction observables can be accommodated using a one-channel distorted-wave
calculation. These calculations go well beyond the DWBA method in that no
Born approximation step is involved. This calculation requires knowledge
of an effective deuteron distorted-wave, being the first component of
the expansion of the $p+n+A$ scattering wave function $\Psi(\vecr,
\vecR)$ in Weinberg states. This component includes accurately breakup
contributions from the small $n$-$p$ separations that dominate the $(d,p)$
reaction amplitude. It is defined as the projection of $\Psi(\vecr,
\vecR)$ onto the transfer reaction vertex, i.e., $V_{np}\mid\phi_d\rangle$.

Johnson and Tandy \cite{Johnson-NPA-1974} showed that, by neglecting couplings
between components in the Weinberg expansion of the three-body wave function,
one obtains a simple prescription for a potential that generates directly (an
approximation to) the first Weinberg component. $(d,p)$ reaction calculations
based on this approximation, known as the adiabatic distorted wave
approximation
(ADWA), have had some success in the analysis of data. Successful and more
complete calculations, that include the couplings between the Weinberg
components
have also been published \cite{Laid-PRC-1993}. We have shown here that there
is a need to develop a simple procedure for correcting the ADWA, focusing
specifically on calculating accurately only the first Weinberg component of
the three-body scattering wave function $\Psi(\vecr,\vecR)$.
This would be especially important for incident energies
of $3-10$ MeV per
nucleon, typical of TRIUMF \cite{Bricault-TRIUMF}, HRIBF at ORNL
\cite{Beene-JPG-2011} (where the
$^{132}$Sn($d,p$)$^{133}$Sn reaction has been measured
\cite{Jones-nature-2010,Jones-PRC-2011}), and ISOLDE \cite{Kester-ISOLDE}, for
which
the influence of closed channels does not allow us to generate reliably this
component using the scheme described above.

\section*{Acknowledgments}

DYP appreciates the warm hospitality he received during his visits to the
University of Surrey. This work is supported by the National Natural Science
Foundation of China under Grants No. 11275018, No. 11021504, and No. 11035001,
and
a project sponsored by the Scientific Research Foundation for Returned
Overseas Chinese Scholars, State Education Ministry. NKT, RCJ, and JAT
gratefully acknowledge the support of the United Kingdom Science and Technology
Facilities Council (STFC) through research Grant No. ST/J000051.

\end{document}